# Strong Spin-Orbit Torque Induced by the Intrinsic Spin Hall Effect in Cr$_{1-x}$Pt$_x$

Qianbiao Liu[1], Jingwei Li[2], Lujun Zhu[3], Xin Lin[1], Xinyue Xie[3], Lijun Zhu[1,4*]

1. *State Key Laboratory of Superlattices and Microstructures, Institute of Semiconductors, Chinese Academy of Sciences, Beijing 100083, China*
2. *Multi-scale Porous Materials Center, Institute of Advanced Interdisciplinary Studies & School of Chemistry and Chemical Engineering, Chongqing University, Chongqing 400044, China*
3. *College of Physics and Information Technology, Shaanxi Normal University, Xi'an 710062, China*
4. *College of Materials Science and Opto-Electronic Technology, University of Chinese Academy of Sciences, Beijing 100049, China*
*\*ljzhu@semi.ac.cn*

We report on a spin-orbit torque study of the spin current generation in Cr$_{1-x}$Pt$_x$ alloy, using the light 3$d$ ferromagnetic Co as the spin current detector. We find that the dampinglike spin-orbit torque of Cr$_{1-x}$Pt$_x$/Co bilayers can be enhanced by tuning the Cr concentration in the Cr$_{1-x}$Pt$_x$ layer, with a maximal value of 0.31 at the optimal composition of Cr$_{0.2}$Pt$_{0.8}$. The mechanism and the efficiency of spin current generation in the Cr$_{1-x}$Pt$_x$ alloy can be fully understood by the characteristic trade-off between the intrinsic spin Hall conductivity of Pt and the spin carrier lifetime in the dirty limit. This suggests that Cr is simply as effective as other metals, oxides, and nitrides (e.g. Hf, Au, Pd, Cu, Ti, MgO, and Si$_3$N$_4$) in enhancing the dampinglike spin-orbit torque generated by the spin Hall effect of a Pt host via strengthening the spin carrier scattering and that alloying Pt with Cr does not employ observable spin current generation via additional spin Hall effect, orbital Hall effect, or interfacial spin-orbit coupling effects. This work also establishes the low-resistivity Cr$_{0.2}$Pt$_{0.8}$ as an energy-efficient spin-orbit torque provider for magnetic memory and computing technologies.

## I. INTRODUCTION

Strong spin-orbit torques (SOTs) have potential to efficiently manipulate magnetization in nonvolatile memory and computing technologies [1-9]. In the simple case of heavy metal/light 3$d$ ferromagnet metal (HM/FM) in which the spin Hall effect (SHE) of the HM is the dominant source of the spin current and the spin current diffused into the FM relaxes mainly via the exchange interaction with the magnetization, the dampinglike SOT efficiency ($\xi_{\text{DL}}^j$) is approximately the product of the spin Hall ratio ($\theta_{\text{SH}}$) of the HM and the spin transparency ($T_{\text{int}}$) of the HM/FM interface, *i.e.*, $\xi_{\text{DL}}^j \approx T_{\text{int}}\theta_{\text{SH}}$. Consequently, $\xi_{\text{DL}}^j$ can be enhanced if $\theta_{\text{SH}}$ of the HM [10-14] or $T_{\text{int}}$ of the HM/FM interface [15-19] can be increased.

Platinum (Pt) is an excellent platform for developing strong spin Hall metals because of the giant intrinsic spin Hall conductivity ($\sigma_{\text{SH}}$) of $\approx 1.6 \times 10^6$ ($\hbar/2e$) $\Omega^{-1}$m$^{-1}$ in the clean limit [16, 20], high electrical conductivity ($\sigma_{xx}$), high spin-mixing conductance ($G_{\text{eff}}^{\uparrow\downarrow}$) of $1.3 \times 10^{15}$ $\Omega^{-1}$m$^{-2}$ in contact with FM [18,19] and ferrimagnetic metals [21]. Theories [20] and experiments [22,23] have well established that $\theta_{\text{SH}} = (2e/\hbar)\sigma_{\text{SH}}/\sigma_{xx}$ of a Pt host can be significantly enhanced by optimizing the trade-off between the intrinsic $\sigma_{\text{SH}}$ and the spin carrier lifetime (which is proportional to $\sigma_{xx}$). In agreement with this mechanism, the scaling of $\sigma_{\text{SH}}$ with $\sigma_{xx}$ coincides for various Pt-based alloys and multilayers [19,24].

However, alloying Pt with Cr has been recently reported to be considerably more [25] or substantially less [26] effective in enhancing $\xi_{\text{DL}}^j$ than alloys and multilayers of Pt and other metals (e.g., Hf [10], Au [11], Pd [12], Cu [9,14], and Ti [21]), oxides (MgO [13]), or nitrides (Si$_3$N$_4$ [27]). It is also surprising that the claimed highest value of $\xi_{\text{DL}}^j$ is one order of magnitude greater for Cr$_{1-x}$Pt$_x$/Co ($\xi_{\text{DL}}^j = 0.92\pm0.07$) in Ref. [25] than for Cr$_{1-x}$Pt$_x$/Pt 0.4/Co ($\xi_{\text{DL}}^j \approx 0.1$) in Ref. [26] despite the same spin-current generating material Cr$_{1-x}$Pt$_x$ and the same loop shift measurement technique. It is particularly motivating to establish in-depth understanding of the spin current physics of the Cr$_{1-x}$Pt$_x$ alloy and to clarify the discrepancies in the literature since Cr is a light metal that has been reported to have rich manifestations such as negative [28,29], positive $\theta_{\text{SH}}$ [30], orbital Hall effect [31,32], and antiferromagnetism [29] in its film form.

In this work, we examine the spin current generation in Cr$_{1-x}$Pt$_x$ alloy using the light 3$d$ ferromagnetic Co as the spin current detector. We find that $\xi_{\text{DL}}^j$ for Cr$_{1-x}$Pt$_x$/Co bilayers can be enhanced to 0.31 by tuning the composition of the Cr$_{1-x}$Pt$_x$ layer. The mechanism and the efficiency of spin current generation in the Cr$_{1-x}$Pt$_x$ alloy can be fully understood by the characteristic trade-off between the intrinsic spin Hall conductivity of Pt and the carrier lifetime in the dirty limit. This reveals that Cr is simply as effective as other metals and oxides (e.g. Au, Pd, Cu, Ti, MgO, and Si$_3$N$_4$)[9-14,21-23,27] in enhancing $\xi_{\text{DL}}^j$ via strengthening the spin carrier scattering in the Pt host.

## II. RESULTS AND DISCUSSIONS

### A. Samples and characterizations

The samples for this work include in-plane magnetic anisotropy (IMA) bilayers of Cr$_{1-x}$Pt$_x$ 5/Co 1.3 and perpendicular magnetic anisotropy (PMA) bilayers of Cr$_{1-x}$Pt$_x$ 5/Co 0.48 (numbers are thicknesses in nm) with different Pt concentration (*x*), as well as IMA Cr$_{1-x}$Pt$_x$ *d*/Co 1.3 with different Cr$_{1-x}$Pt$_x$ thickness (*d*). Each sample is sputter-deposited on a Si/SiO$_2$ substrate at room temperature with a 1 nm Ta adhesion layer and protected by a MgO 1.6/Ta 1.5 bilayer. The 1 nm Ta adhesion layer is highly resistive and contributes negligible spin current into the Co. The top Ta capping layer is fully oxidized upon exposure to the atmosphere. Scanning transmission electron microscopy image and electron-dispersive x-ray spectrum mapping results in Fig. 1(a) reveal good elemental uniformity of Cr and Pt within the Cr$_{1-x}$Pt$_x$ layer, polycrystalline structure with large lateral grain size of 15 nm, and reasonably sharp interface with the Co layer. High-resolution STEM images of the interface regions (Fig. 1(b)) indicate face-centered cubic (fcc) structure for both Co and Cr$_{1-x}$Pt$_x$ layers and epitaxial growth of Co layer on top of the Cr$_{1-x}$Pt$_x$ along (111) orientation. This is consistent with previous reports that Cr$_{1-x}$Pt$_x$ alloy prepared at room temperature crystallizes into the chemically disordered fcc structure in the composition range of $x \geq 0.4$ [33]. Atomic force microscopy measurements (Fig. 1(c)) indicate that our Cr$_{1-x}$Pt$_x$ layers have fairly smooth surfaces in a large scale, for instance, with a small square root surface roughness of 0.3 nm for the Cr$_{1-x}$Pt$_x$ with $x = 0.2$. X-ray diffraction $\theta$-$2\theta$ patterns show only the fcc (111) peaks for all the Cr$_{1-x}$Pt$_x$ layers studied in this work (Fig. 1(d)), reaffirming that the Cr$_{1-x}$Pt$_x$ forms a chemically disordered fcc alloy with preferred (111) orientation. From the Cr$_{1-x}$Pt$_x$ (111) peak, which shifts increasingly more to a higher $2\theta$ angle with increasing Cr concentration, we determine that the average lattice constant of the Cr$_{1-x}$Pt$_x$ is a linear function of the Pt concentration (Fig. 1(e)), in good agreement with the Vegard's law.

Superconducting quantum interference device measurements reveal that the saturation magnetization ($M_s$) is $\approx 1370$ emu/cm$^3$ for the 1.3 nm Co, $\approx 1210$ emu/cm$^3$ for the 0.48 nm Co. The dependence on the Co thickness of magnetic moment per area suggests negligible dead layer for these Co layers (Fig. A1 in the Appendix). The Cr$_{1-x}$Pt$_x$ layers are paramagnetic ($M_s \approx 0$ emu/cm$^3$) and show no measurable magnetization or anomalous Hall effect (Fig. A1 in the Appendix), which is consistent with the literature reports [33,34]. The samples were then patterned by photolithography and ion milling into 5×60 µm$^2$ Hall bars followed by deposition of 5 nm Ti and 150 nm Pt as contacts for electrical measurements. The resistivity ($\rho_{xx}$) for Cr$_{1-x}$Pt$_x$ is determined by measuring the resistivity of control Cr$_{1-x}$Pt$_x$ single layers and by subtracting the conductance of Ta 1/Co 0.48 (1.3)/MgO 1.6/Ta 1.6 from the whole stack Ta 1/Cr$_{1-x}$Pt$_x$ $d$/Co 0.48 (1.3)/MgO 1.6/Ta, respectively. As shown in Fig. 1(f), the values of $\rho_{xx}$ determined from the two methods consistently vary from 37 µΩ cm for $x = 1$ to 135 µΩ cm for $x = 0.5$. Figure 1(g) shows the dependence on the thickness of the values of $\rho_{xx}$ for the Cr$_{0.2}$Pt$_{0.8}$ as determined by averaging the values estimated from the two methods.

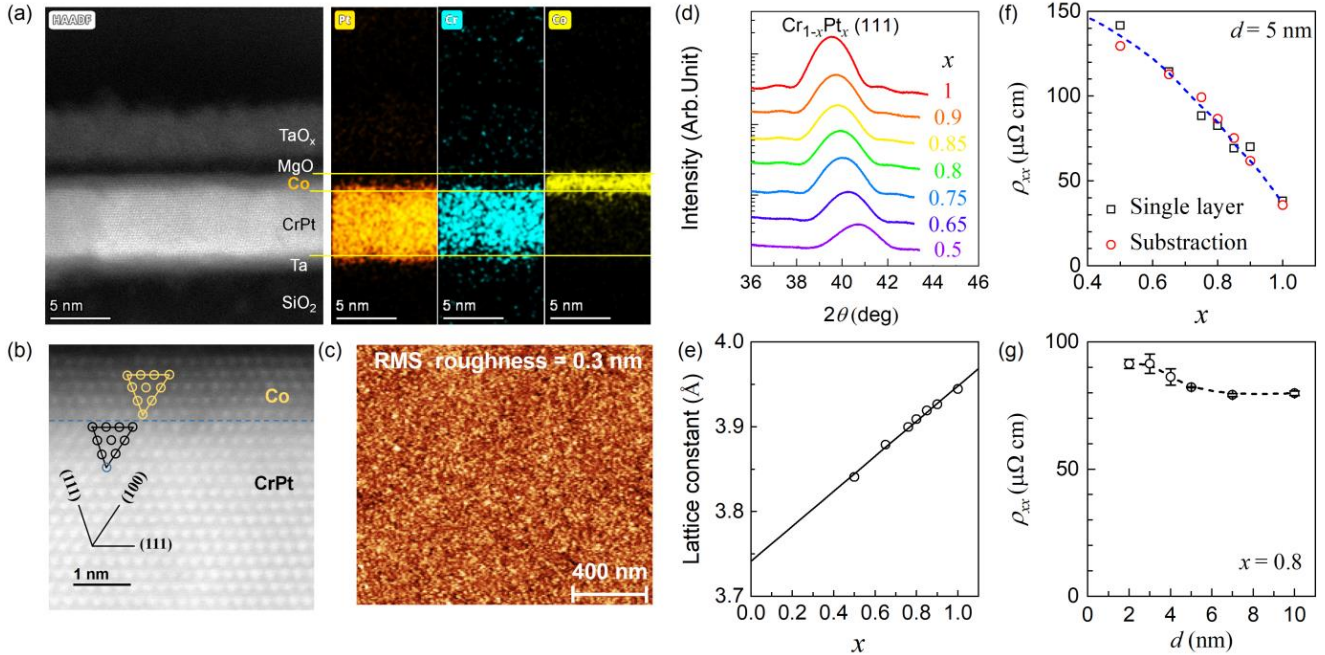

**Fig. 1** (a) Cross-sectional scanning transmission electron microscopy image and electron dispersive x-ray spectrum mapping of Pt, Cr, and Co for a Cr$_{0.2}$Pt$_{0.8}$ 5/Co 1.3 sample. (b) High-resolution scanning transmission electron microscopy image of the interface of the Cr$_{0.2}$Pt$_{0.8}$ 5/Co 1.3 bilayer, suggesting the epitaxial growth of Co on Cr$_{1-x}$Pt$_x$. (c) Atomic force microscopy image (2×2 µm$^2$ in area) of the Cr$_{0.2}$Pt$_{0.8}$ 5/Co 1.3 sample, indicating a small square root roughness of 0.3 nm. (d) X-ray diffraction patterns of the Cr$_{1-x}$Pt$_x$ 5/Co 1.3 with different Cr$_{1-x}$Pt$_x$ compositions. (e) Lattice constant of Cr$_{1-x}$Pt$_x$ plotted as a function of $x$. (f) $x$ dependence of $\rho_{xx}$ for the Cr$_{1-x}$Pt$_x$ as determined by measuring the resistivity of control Cr$_{1-x}$Pt$_x$ single layers (black squares) and by subtracting the conductance of control stacks Ta 1/Co 0.48, 1.3/MgO 1.6/Ta 1.6 from the whole stack Ta 1/ Cr$_{1-x}$Pt$_x$ $d$/Co 0.48, 1.3/MgO 1.6/Ta (red circles). (g) The averaged resistivity for the Cr$_{1-x}$Pt$_x$ with different thicknesses. Dashed lines in (c), (g) and (h) are to guide the eyes.

## B. Composition dependence of spin-orbit torques

The dampinglike SOT efficiencies of $Cr_{1-x}Pt_x$ 5/Co 1.3 bilayers ($x \geq 0.5$) are first determined by angle-dependent "in-plane" harmonic Hall voltage response (HHVR) measurements [11,35]. While applying a sinusoidal electric field ($E$) of 33.3 kV/m onto the bar orientated along the $x$ axis, the in-phase first and out-of-phase second HHVRs, $V_{1\omega}$ and $V_{2\omega}$, are collected as a function of the angle ($\varphi$) of the in-plane magnetic field ($H_{xy}$) with respect to the current under different magnetic field magnitudes (1250-3250 Oe). Within the macrospin approximation,

$$V_{2\omega} = V_{DL+ANE}\sin\varphi + V_{FL+Oe}\sin\varphi\cos2\varphi, \quad (1)$$

where $V_{DL+ANE} = -V_{AHE}H_{DL}/2(H_{xy}+|H_k|) + V_{ANE}$, $V_{FL+Oe} = -V_{PHE}(H_{FL} + H_{Oe})/2H_{xy}$, $V_{AHE}$ is the anomalous Hall voltage, $V_{PHE}$ the planar Hall voltage, and $V_{ANE}$ the anomalous Nernst voltage induced by the vertical thermal gradient, $H_k$ the perpendicular anisotropy field. The values of $V_{AHE}$ and $H_k$ are determined from the dependence of $V_{1\omega}$ on the swept out-of-plane field ($H_z$) under zero $H_{xy}$ (Fig. 2(a)). The $V_{DL+ANE}$ of the $Cr_{1-x}Pt_x$ d/Co 1.3 for each magnitude of $H_{xy}$ is extracted from fit of the $\varphi$ dependence of $V_{2\omega}$ to Eq. (1). (See Fig. 2(b) for two representative examples for the $Cr_{0.2}Pt_{0.8}$ 5/Co 1.3). The slope and the intercept of the linear fit of $V_{DL+ANE}$ vs $-V_{AHE}/2(H_{xy}+|H_k|)$ give the values of $H_{DL}$ and $V_{ANE}$, respectively (Fig. 2(c)). Figure 3(a) shows $\xi_{DL}^j$ as calculated following the relation

$$\xi_{DL}^j = (2e/\hbar)\mu_0 M_s t_{Co} H_{DL}\rho_{xx}/E, \quad (2)$$

with $e$, $\mu_0$, $t_{Co}$, and $\hbar$ being the elementary charge, the permeability of vacuum, the Co thickness, the reduced Planck's constant, respectively. $\xi_{DL}^j$ for the IMA $Cr_{1-x}Pt_x$ 5/Co 1.3 firstly increases from 0.18 at $x = 0$ to the peak value of 0.31 at $x = 0.80$, and then gradually drops to 0.20 at $x = 0.5$. We find a strong anisotropy in the film plane in the $Cr_{1-x}Pt_x$ 5/Co 1.3 with $x = 0$ and 0.2, preventing analysis of $\xi_{DL}^j$ at those compositions (see Fig. A2 in the Appendix).

We find a similar composition dependence of $\xi_{DL}^j$ for the PMA $Cr_{1-x}Pt_x$ 5/Co 0.48 bilayers from out-of-plane HHVR measurements [36]. As indicated by the fairly square Hall voltage hysteresis in Fig. 2(d), the $Cr_{1-x}Pt_x$ 5/Co 0.48 with $x \geq 0.75$ show good PMA, which allows us to perform high-quality out-of-plane HHVR analyses. As we show in Figs. 2(e) and 2(f), $V_{1\omega}$ and $V_{2\omega}$ for the $Cr_{1-x}Pt_x$ 5/Co 0.48 with $x \geq 0.75$ are well-defined parabolic and linear functions of the in-plane magnetic field $H_x$, respectively, from which the dampinglike SOT field is determined as $H_{DL} = -2(\partial V_{2\omega}/\partial H_x)/(\partial^2 V_{1\omega}/\partial H_x^2)$. Note that there is only negligible anomalous Nernst effect [37,38] in our $Cr_{1-x}Pt_x$ 5/Co 0.48 bilayers (Fig. A3 in the Appendix). We have ignored the so-called "planar Hall correction" which is generally found to cause errors when not negligible [12,39,40] (also see Fig. A4 in the Appendix). As we show in Fig. 3(a), $\xi_{DL}^j$ for the $Cr_{1-x}Pt_x$ 5/Co 0.48 also peaks at $x \approx 0.80$. Interestingly, the values of $\xi_{DL}^j$ for the $Cr_{1-x}Pt_x$ 5/Co 0.48 are slightly smaller than that for the $Cr_{1-x}Pt_x$ 5/Co 0.48. This is likely because the

Co thickness of 0.48 nm is smaller than the spin dephasing length of the Co such that a portion of spin current approaches and gets reflected at the top Co/MgO interface. It is expected that the polarization of such reflected spin current is rotated relative to the incident spin current [41,42], which leads to decrease of the effective SOT exerted on the Co layer. The difference in the torque efficiencies for the PMA $Cr_{1-x}Pt_x$ 5/Co 0.48 and the IMA $Cr_{1-x}Pt_x$ 5/Co 1.3 cannot be attributed to spin memory loss because the interfacial magnetic anisotropy energy density ($K_s^{ISOC}$) is smaller at the $Cr_{0.2}Pt_{0.8}$ 5/Co 0.48 interface than the $Cr_{0.2}Pt_{0.8}$ 5/Co 1.3 interface (Fig. A6 in Appendix). The PMA of the $Cr_{1-x}Pt_x$ 5/Co 0.48 becomes weak for $x < 0.75$ (Fig. A5 in the Appendix), preventing out-of-plane HHVR analysis.

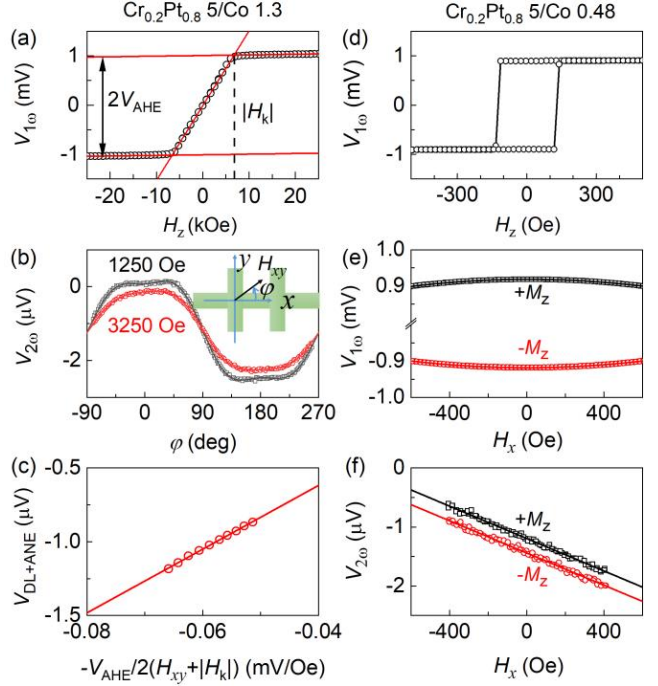

**Fig. 2** (a) $V_{1\omega}$ vs $H_z$. (b) $V_{2\omega}$ vs $\varphi$, and (c) $V_{DL+ANE}$ vs $-V_{AHE}H_{DL}/2(H_{xy}+|H_k|)$ for the $Cr_{0.2}Pt_{0.8}$ 5/Co 1.3. (d) $V_{1\omega}$ vs $H_z$, (e) $V_{1\omega}$ vs $H_x$, and (f) $V_{2\omega}$ vs $H_x$ for the $Cr_{0.2}Pt_{0.8}$ 5/Co 0.48. Solid lines in (a), (c), and (f) represent linear fits; the solid curves represent fits of the data to Eq. (1) in (b) and to a parabolic function in (e). The inset in (b) shows the measurement coordinate.

## C. Mechanism of spin-orbit torque

We now show that the spin current generation in the $Cr_{1-x}Pt_x$ is dominated by the intrinsic SHE. As plotted in Figs. 3(b), $\xi_{DL}^j$ and the apparent spin Hall conductivity $T_{int}\sigma_{SH} = (\hbar/2e)\xi_{DL}^j/\rho_{xx}$ of the $Cr_{0.2}Pt_{0.8}$ d/Co 1.3 are large constants at large $d$ values and reduces gradually towards zero as $d$ decreases, which is consistent with the SHE being the dominant mechanism of the spin current generation in the $Cr_{1-x}Pt_x$. A more unambiguous piece of evidence for the intrinsic SHE is the characteristic decrease of $T_{int}\sigma_{SH}$ of $Cr_{1-x}Pt_x$ 5/Co 1.3 with $\sigma_{xx}$ in the dirty limit. Here, $T_{int}$ is set by spin backflow [16,19,43,44] and spin memory loss [17], i.e., $T_{int} \approx T_{int}^{SML}T_{int}^{SBF}$ [23,24]. Drift-diffusion model [16,43,44] predicts

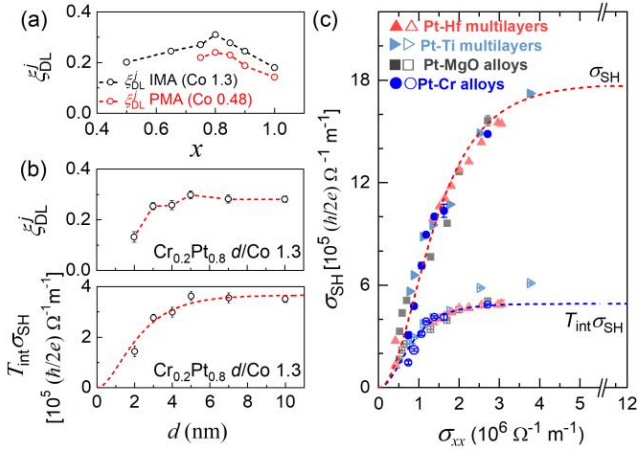

**Fig. 3** (a) $\xi_{DL}^j$ for $Cr_{1-x}Pt_x$ 5/Co 1.3 (IMA) and $Cr_{1-x}Pt_x$ 5/Co 0.48 (PMA) with different Pt concentration. (b) $\xi_{DL}^j$ and $T_{int}\sigma_{SH}$ for $Cr_{0.2}Pt_{0.8}$ $d$/Co 1.3 with different thickness $d$. (c) Spin Hall conductivities $T_{int}\sigma_{SH}$ and $\sigma_{SH}$ for $Cr_{1-x}Pt_x$ 5/Co 1.3, $Pt_x(MgO)_x$/Co [13], Pt-Hf multilayers/Co [22], and Pt-Ti multilayers/Co [23]. The lines are for guidance of eyes.

$$T_{int}^{SBF} = (1-\text{sech}(d/\lambda_s))(1+\tanh(d/\lambda_s)/2\lambda_s\rho_{xx}G_{eff}^{\uparrow\downarrow}), \quad (3)$$

where $1/\lambda_s\rho_{xx} \approx 1.3 \times 10^{15}$ $\Omega^{-1}m^{-2}$ for Pt [45,46]. Our previous study shows that $T_{int}^{SML} \approx 1-0.23K_s^{ISOC}$ for IMA Pt/Co, with $K_s^{ISOC}$ in erg/cm$^2$. The values of $K_s^{ISOC}$ for the interfaces of $Cr_{1-x}Pt_x$ 5/Co 1.3 are estimated by subtracting the interfacial PMA energy density of $\approx 0.6$ erg/cm$^2$ [17] from the total interfacial PMA energy density ($K_s$) of the two interfaces of the $Cr_{1-x}Pt_x$ 5/Co 1.3/MgO (Fig. A7 in the Appendix). $K_s$ is estimated following the relation $H_k \approx -4\pi M_s + 2K_s/M_st_{Co}$.

In Fig. 3(c), we compare the values of $T_{int}\sigma_{SH}$ and $\sigma_{SH}$ of the $Cr_{1-x}Pt_x$ 5/Co 1.3 with those of $Pt_x(MgO)_x$/Co [13], Pt-Hf multilayers/Co [23] and Pt-Ti multilayers/Co [24] as a function of $\sigma_{xx}$. Clearly, both $T_{int}\sigma_{SH}$ and $\sigma_{SH}$ for the $Cr_{1-x}Pt_x$ follow the characteristic decrease of the *intrinsic* spin Hall conductivity with $\sigma_{xx}$ in the dirty limit, and overall coincide with those of other systems in which the intrinsic SHE is the dominant source of the spin current. This is striking evidence that the dominant mechanism of the spin current generation in the $Cr_{1-x}Pt_x$ is the intrinsic SHE and that there is minimal additional spin current generation via any other mechanisms (e.g. new spin Hall effect or orbital Hall effect from Cr or interfacial SOC effect) in this material. This is consistent with previous experimental [11-13] and theoretical [47] reports of absence of any important extrinsic spin Hall effect in Pt-rich alloys. We find that the characteristic variations of $T_{int}\sigma_{SH}$ and $\sigma_{SH}$ with the longitudinal conductance for the Pt alloys and multilayers can be empirically fit by

$$T_{int}\sigma_{SH} = T_{int}\sigma_{SH0}(1-\text{sech}(\sigma_{xx}/\sigma_{xx0})), \quad (4)$$

with the clean-limit apparent spin Hall conductivity $T_{int}\sigma_{SH0}$ of $\approx 4.95\times10^5$ $(\hbar/2e)\Omega^{-1}$ m$^{-2}$ and with the characteristic conductivity $\sigma_{xx0}$ of $\approx 5.0\times10^5$ $\Omega^{-1}$ m$^{-1}$. We also find that the fieldlike SOT of these samples are much smaller than the dampinglike torque (Fig. A4 in the Appendix), which is consistent with the intrinsic SHE in $Cr_{1-x}Pt_x$ alloys being the dominant source of the spin current associated with the SOTs.

In principle, interface-generated spin currents, if significant, can add to or subtract from the spin Hall spin current of HM/FM samples and affect the measured strength of SOTs. However, this usually seems to be a minor effect at magnetic interfaces. A recent SOT experiment [48] has established clean evidence that, under the same electric field, the efficiency of spin current generation by interfacial SOC effect (e.g. Rashba-Edelstein(like) effect, inverse spin galvanic effect, or spin filtering effect) should be at least 2-3 orders of magnitude smaller than that of Pt even when the interfacial SOC is quite strong (e.g. at Ti/FeCoB interfaces). The absence of a significant interfacial spin current generation is also well supported by: (i) a spin Seebeck/ISHE experiment that spin current generation is absent at $Bi/Ag/Y_3Fe_5O_{12}$ and $Bi/Y_3Fe_5O_{12}$ interfaces [49,50] prepared by different techniques; (ii) the universal observation of strong scaling of the SOT efficiencies or the effective spin Hall ratio with the resistivity and the layer thickness of HMs [13,14,22,23,45,46,51,52] (see Figs. 3(b) and 3(c) for the $Cr_{1-x}Pt_x$/Co case), topological insulators [53], nonmagnetic complex oxides [54]; (iii) the clear dependence of inverse spin Hall voltage on the spin-mixing conductance of the interfaces of magnetic oxides (e.g. $Y_3Fe_5O_{12}$) in spin Seebeck [49,50,55] and spin pumping [56]; and (iv) absence of SOT enhancement with increasing interfacial SOC strength [17,48,57] (see Fig. 3(a) and Fig. A7 in the Appendix for the $Cr_{1-x}Pt_x$/Co). This is consistent with the theories [16,58-60] that interfacial SOC has negligible contribution to the dampinglike SOT via the two-dimensional Rashba-Edelstein(like) effect of magnetic interfaces.

Here, we note that a previous loop shift study by Quan *et al.* [26] only reported a maximum $\xi_{DL}^j$ of 0.1 for $Cr_{1-x}Pt_x$ 8/Pt 0.4/Co 0.9, however, those samples were annealed at 370 °C and thus likely have very strong spin memory loss at the Pt/Co interface [17]. Meanwhile, at the optimal of that work, the resistivity $Cr_{1-x}Pt_x$ composition was only 56 μΩ cm and the Co magnetization was as low as 820 emu/cm$^3$. $\xi_{DL}^j$ of a HM/FM is positively correlated to the resistivity of the HM in this range, and a low magnetization may reduce the spin-dephasing length to be below the layer thickness of the FM and lower $\xi_{DL}^j$ due to the resultant polarization rotation [41,42] of the reflected spin current at the top interface. We have also noticed that another loop shift experiment [25] claimed $\xi_{DL}^j$ of 0.2-0.34 for Pt/Co and 0.9-0.94 for $Cr_{1-x}Pt_x$/Co in the composition range of $0.5 \leq x \leq 0.7$, which are surprisingly large compared to our expectation from the existing studies on Pt [9-14,19,22-24]. However, for reasons unknown, that work estimated resistivities that are a factor of 2-3 greater than ours in their calculation of $\xi_{DL}^j$ and their $Cr_{1-x}Pt_x$/Co samples also remained PMA in the wide concentration range of $x = 0.5-1$ even when $t_{Co}$ was as large as 1.6 nm. The latter is in contrast to our observation that the $Cr_{1-x}Pt_x$/Co becomes in-plane magnetized when $x \leq 0.65$ regardless of $t_{Co}$ or when $t_{Co}$ is 1.3 nm regardless of the Pt concentration. Note that the $Cr_{1-x}Pt_x$ 8/Pt 0.4/Co 0.9 bilayers in Ref. 25 also required post-growth annealing at 370 °C to obtain PMA for the thin Co in the composition range of (0.5 ≤ $x$ ≤1). Finally, we do not attribute the large discrepancies of the $\xi_{DL}^j$ values in this work, Ref. 25, and Ref. 26 to the measurement techniques since loop shift and HHVR techniques can yield consistent SOT results in other material systems (see detailed discussions in [61]).

### D. Practical impact

Finally, we note that $Cr_{0.2}Pt_{0.8}$, which has $\rho_{xx}$ ≈ 80 μΩ cm, $\lambda_s$ ≈1.6 nm, $\theta_{SH}$ ≈ 0.7, is a compelling spin Hall metal for technological applications. As shown in Fig. 4(a), a 5 nm $Cr_{0.2}Pt_{0.8}$ can switch a Co layer with high PMA and coercivity ($H_k$ ≈ 4900 Oe, $H_c$ ≈ 120 Oe) at a current density of $1.2 \times 10^7$ A/cm$^2$. More impressively, as we show in Fig. 4(b), magnetization switching of strong PMA $Cr_{0.2}Pt_{0.8}$ 5/Ti 0.8/CoFeB 1 nm ($H_k$ ≈ 2500 Oe, $H_c$ ≈ 100 Oe) can be achieved at a current density as low as ≈ $4.8 \times 10^6$ A/cm$^2$ in the $Cr_{0.2}Pt_{0.8}$ layer. As compared in Fig. 4(c), $Cr_{0.2}Pt_{0.8}$ has high $\xi_{DL}^j$ and relatively low $\rho_{xx}$ that are similar to $Pt_{0.6}(MgO)_{0.4}$ [13], $Pt_{0.7}(Si_3N_4)_{0.3}$ [27], $Au_{0.25}Pt_{0.75}$ [11], Pt-Ti multilayers [23]. Interestingly, the values of $\xi_{DL}^j$ and $1/\sigma_{xx} = \rho_{xx}$ for various Pt alloys and multilayers approximately follow

$$\xi_{DL}^j = (2e/\hbar)T_{int}\sigma_{SH0}/\sigma_{xx}(1-\text{sech}(1/\sigma_{xx0})), \quad (5)$$

and *statistically* indicates maximization in the resistivity ($1/\sigma_{xx}$) range of 80-160 μΩ cm. Since $\xi_{DL}^j$ expected from this scaling varies with $\rho_{xx}$ only slightly in this resistivity range, the exact value of the largest $\xi_{DL}^j$ and the corresponding resistivity of different specific material systems may vary due to varying strengths of spin backflow and spin memory loss at the interface and varying thicknesses and spin relaxation rates of the heavy metal and the ferromagnet in different studies. This should explain why the largest $\xi_{DL}^j$ for the $Cr_{1-x}Pt_x$ 5/Co 1.3 (IMA) and the $Cr_{1-x}Pt_x$ 5/Co 0.48 (PMA) with different Pt concentration shows up at the composition $x$ = 0.2 ($\rho_{xx}$ ≈ 80 μΩ cm, Fig. 3(a)) rather than lower $x$ values (more resistive, Fig. 1(f)). Since the switching power of a nanomagnet sit on a spin Hall metal channel (such as SOT-MTJs) can be estimated as

$$P \propto [(1+t_{FM}\sigma_{FM}/d\sigma_{xx})/\xi_{DL}^j]^2\rho_{xx}, \quad (6)$$

In Fig. 4(c), we plot the normalized power consumption $P$ of a FeCoB nanomagnet ($t_{FM}$ = 1.6 nm, $1/\sigma_{FM}$= 135 μΩ cm) sit on different strong spin Hall metal channels as calculated using the measured values of $1/\sigma_{xx}$ and $\xi_{DL}^j$. The power consumption for the $Cr_{0.2}Pt_{0.8}$ device is comparable with the optimized $Pt_{0.6}(MgO)_{0.4}$ [13], $Pt_{0.7}(Si_3N_4)_{0.3}$ [27], $Au_{0.25}Pt_{0.75}$ [11], $Pd_{0.25}Pt_{0.75}$ [12], and Pt-Ti multilayers [23], and represents the lower limit of the Pt-based alloys and multilayers as indicated by the red dashed line that plots the power consumption by using $\xi_{DL}^j$ and $\sigma_{xx}$ predicted by Eq. (5). Therefore, $Cr_{0.2}Pt_{0.8}$ is an excellent spin Hall metal that is comparable to the optimized Pt alloys and advantageous over $\beta$-Ta [1], $\beta$-W [3], clean-limit Pt [62], in terms of energy efficiency, impedance, and endurance when used in spin-orbit technologies.

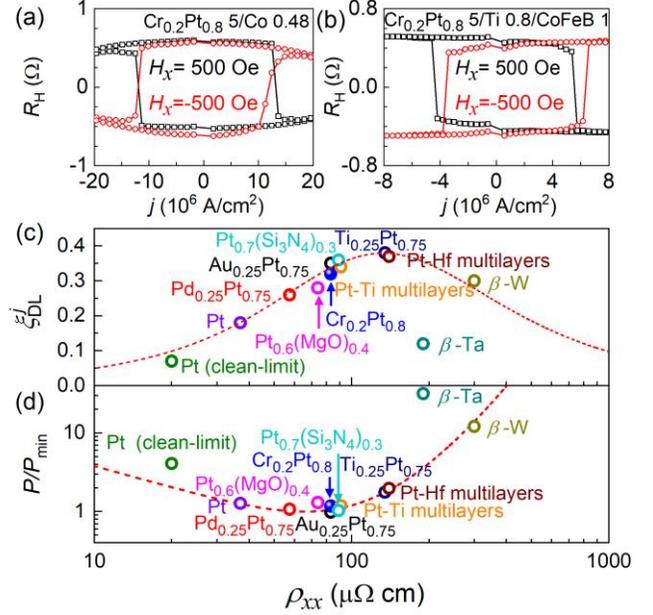

**Fig.** 4 Hall resistance vs in-plane current density in the $Cr_{0.2}Pt_{0.8}$ layer for (a) the $Cr_{0.2}Pt_{0.8}$ 5/Co 0.48 and (b) the $Cr_{0.2}Pt_{0.8}$ 5/Ti 0.8/CoFeB 1 under in-plane field of ±500 Oe. (c) $\xi_{DL}^j$ and $\rho_{xx}$ for spin Hall metal/FM bilayers. The spin Hall metals are Pt, $Cr_{0.2}Pt_{0.8}$, $Ti_{0.25}Pt_{0.75}$ [21], $Pt_{0.6}(MgO)_{0.4}$ [13], $Au_{0.25}Pt_{0.75}$ [11], $Ti_{0.25}Pt_{0.75}$ [11], $Pt_{0.7}(Si_3N_4)_{0.3}$ [27], $Pd_{0.25}Pt_{0.75}$ [12], $\beta$-Ta [1], $\beta$-W [3], and clean-limit Pt [62]. The dashed line in (c) represents the plot of Eq. (5), which indicates maximization of $\xi_{DL}^j$ for Pt-based HM/FM systems in the resistivity range of 80-160 μΩ cm. (d) Normalized power consumption of SOT-MTJs based on different spin Hall metals as estimated using Eq. (6). The dashed line in (d) plots the power consumption expected for a spin Hall metal with $\xi_{DL}^j$ and $\rho_{xx}$ following Eq. (5). $P_{min}$ is the minimum value of power predicted by the dashed line.

### III. CONCLUSIONS

In conclusion, we have presented that the dampinglike spin-orbit torque efficiency of $Cr_{1-x}Pt_x$/Co bilayers can be enhanced by tuning the Cr concentration in the $Cr_{1-x}Pt_x$ layer, with a maximal value of 0.31 at the optimal composition of $Cr_{0.2}Pt_{0.8}$. The efficiency of the spin current generation in the $Cr_{1-x}Pt_x$ alloy is found to be dominated by the characteristic trade-off between the intrinsic spin Hall conductivity of Pt and the spin carrier lifetime in the dirty limit and coincide well with those of other Pt-based spin Hall metal systems. Cr turns out to be simply as effective as other metals, oxides, and nitrides (e.g. Hf, Au, Pd, Cu, Ti, MgO, and $Si_3N_4$) in enhancing the dampinglike spin-orbit torque generated by the spin Hall effect of a Pt host. We find no obvious evidence for spin current generation by a mechanism other than the intrinsic SHE of Pt in $Cr_{1-x}Pt_x$. This work establishes the low-resistivity $Cr_{0.2}Pt_{0.8}$ as an energy-efficient SOT provider for magnetic memory and computing applications.

## ACKNOWLEDGMENTS

This work was supported partly by the Strategic Priority Research Program of the Chinese Academy of Sciences (XDB44000000) and by the National Natural Science Foundation of China (Grants No. 12274405 and 51901121).

## APPENDIX

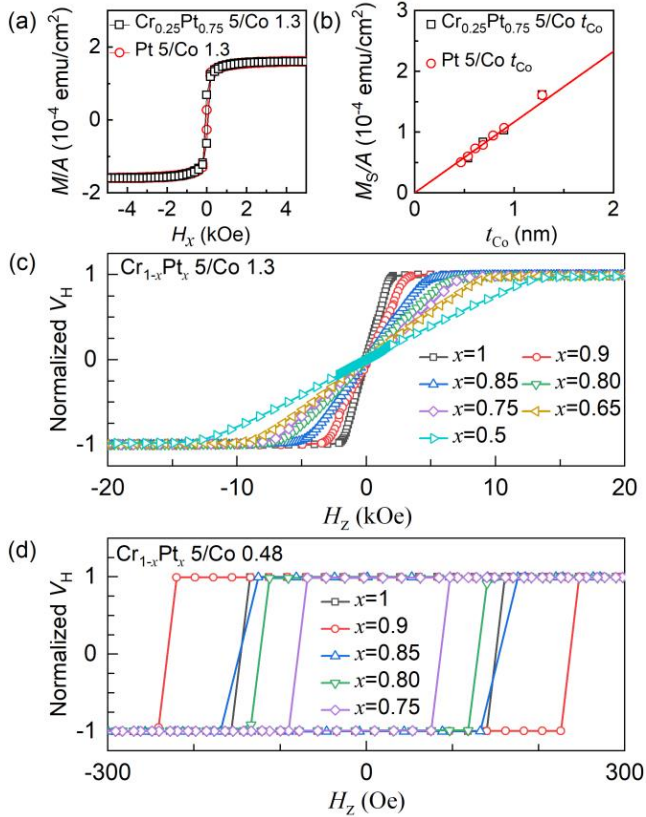

Fig. A1 Absence of magnetic and anomalous Hall voltage contributions from the $Cr_{1-x}Pt_x$ layer. (a) In-plane magnetic moment per area vs in-plane magnetic field for the Pt 5 nm/Co 1.3 nm and for the $Cr_{0.25}Pt_{0.75}$ 5 nm/Co 1.3 nm, suggesting no difference in saturation magnetization for samples with different Pt concentration and thus no magnetic contribution in the $Cr_{0.25}Pt_{0.75}$ layer. (b) The Co thickness dependence of the saturation magnetic moment per area for the $Cr_{1-x}Pt_x$/Co $t_{Co}$ bilayers, extrapolating zero magnetization at zero Co thickness. (c) In-plane non-hysteretic Hall voltage loops for the $Cr_{1-x}Pt_x$ 5 nm/Co 1.3 nm samples. (d) Square Hall voltage loops for the $Cr_{1-x}Pt_x$ 5 nm/Co 0.48 nm samples. Data in (c) and (d) suggests no anomalous Hall voltage contributed by the $Cr_{1-x}Pt_x$ layers in the in-plane and out-of-plane directions. These results are well consistent with expectation that the $Cr_{1-x}Pt_x$ layers are paramagnetic.

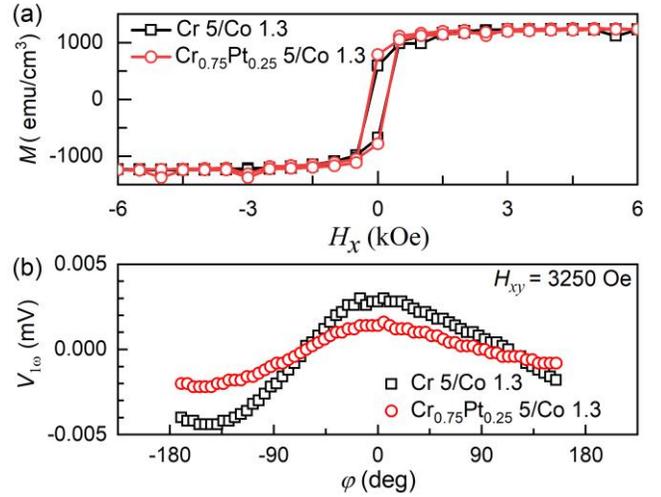

Fig. A2 (a) In-plane magnetization vs in-plane magnetic field and (b) First harmonic Hall voltage response $V_{1\omega}$ vs in-plane angle of field ($\varphi$) with respect to the current (field= 3250 Oe) for Cr 5 nm/Co 1.3 nm and $Cr_{0.75}Pt_{0.25}$ 5 nm/Co 1.3 nm. The very small magnitude and the absence of a $\sin2\varphi$ dependence of $V_{1\omega}$ indicate a strong anisotropy in the film plane that prevents magnetization rotation by the in-plane magnetic field in harmonic Hall voltage response analysis.

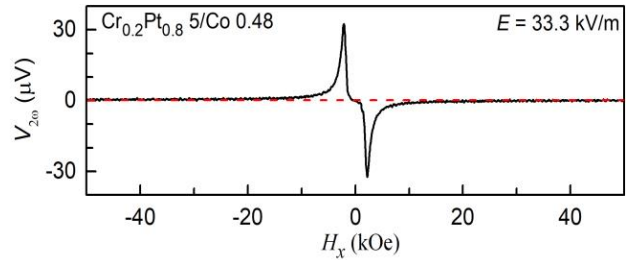

Fig. A3 Second harmonic Hall voltage response ($V_{2\omega}$) vs in-plane magnetic field ($H_x$) for the $Cr_{0.2}Pt_{0.8}$ 5 nm/Co 0.48 nm, suggesting only negligible anomalous Nernst effect.

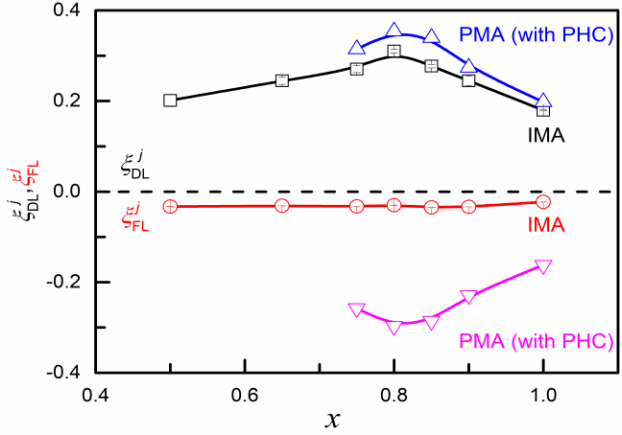

Fig. A4 Dependence on the Pt concentration of the spin-torque efficiencies ($\xi_{DL}^{j}$ and $\xi_{FL}^{j}$) for the $Cr_{1-x}Pt_x$ 5 nm/Co 0.5 nm bilayers as determined from out-of-plane harmonic Hall voltage response (HHVR) measurement with the "planar Hall correction (PHC)" and for $Cr_{1-x}Pt_x$ 5 nm/Co 1.3 nm as determined from in-plane angle-dependent HHVR measurement. The values of $\xi_{DL}^{j}$ and $\xi_{FL}^{j}$ estimated with the planar Hall correction deviate significantly from the in-plane results.

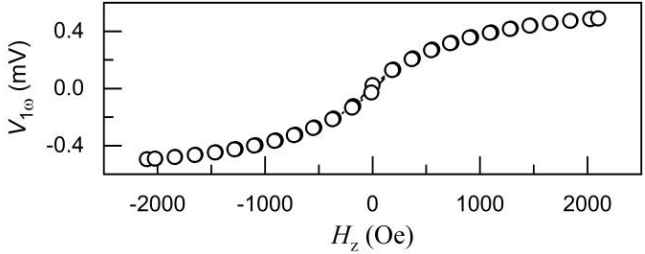

Fig. A5 First harmonic Hall voltage response ($V_{1\omega}$) vs out-of-plane magnetic field ($H_z$) for the $Cr_{0.35}Pt_{0.65}$ 5 nm/Co 0.48 nm, suggesting a perpendicular magnetic anisotropy that is too weak to allow for reliable out-of-plane harmonic Hall voltage response analysis.

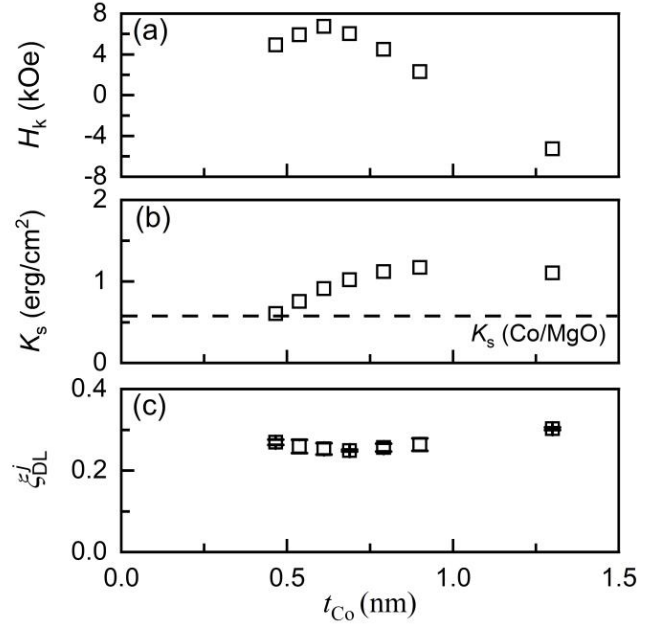

Fig. A6 Dependence on the Co thickness of (a) the perpendicular anisotropy field ($H_k$), (b) the interfacial magnetic anisotropy energy density ($K_s$), and (c) the dampinglike torque efficiency ($\xi_{DL}^{j}$) for the $Cr_{0.2}Pt_{0.8}$/Co $t_{Co}$. $K_s$ is negligible for the $Cr_{0.2}Pt_{0.8}$ 5 nm/Co 0.48 nm but strong for the $Cr_{0.2}Pt_{0.8}$ 5 nm/Co 1.3 nm.

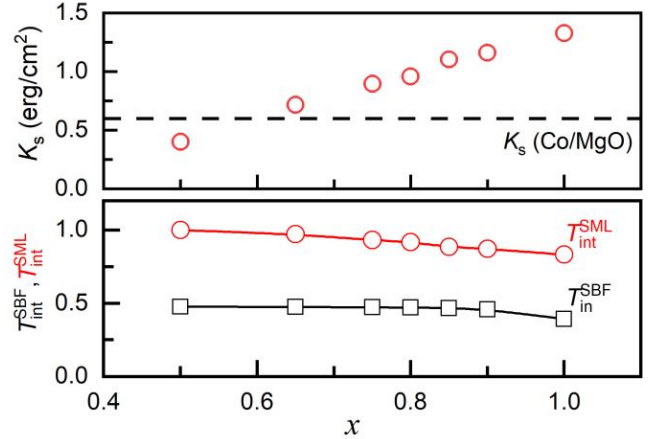

Fig. A7 Dependence on the Pt concentration $x$ of (a) interfacial perpendicular magnetic anisotropy energy density ($K_s$) and (b) spin transparencies set by spin backflow ($T_{int}^{SBF}$, as calculated following Eq. (3) in the main text) and by spin memory loss ($T_{int}^{SML}$, as calculated following the relation $T_{int}^{SML} \approx 1-0.23K_s^{ISOC}$) for the $Cr_{1-x}Pt_x$ 5/Co 1.3/MgO samples. $K_s$ for the $Cr_{0.5}Pt_{0.5}$/Co 1.3/MgO is smaller than that of the Co/MgO interface ($\approx 0.6$ erg/cm$^2$) likely due to a non-negligible in-plane volume anisotropy of Co.